\documentstyle [aps,epsfig,float,twocolumn,prl]{revtex}
\begin{document}
\twocolumn[\hsize\textwidth\columnwidth\hsize\csname @twocolumnfalse\endcsname
\draft
\title{Renormalization of resonant tunneling in MOSFETs }
\author{D. Mozyrsky$^1$ {\rm ,} I. Martin$^1$ {\rm ,}
A. Shnirman$^2$, and M. B. Hastings$^1$\\
$^1$Theoretical Division, Los Alamos National Laboratory, Los Alamos, NM 87545, USA\\
$^2$Institut f\"{u}r Theoretische Festk\"{o}rperphysik, Universit\"{a}t Karlsruhe, D-76128,
Karlsruhe, Germany} \maketitle
\begin{abstract}
We study tunneling between a localized defect state and a
conduction band in the presence of strong electron-electron and
electron-phonons interactions. We derive the tunneling rate as a
function of the position of the defect energy level relative to
the Fermi energy of conduction electrons. We argue that our
results can explain the large tunneling timescales observed in
experiments on random telegraph signals in ${\rm Si}$
metal-oxide-semiconductor field effect transistors.
\end{abstract}

]  It has been long realized that two-state systems can have
profound effects on micro- and nano-scale electronic devices,
leading to random switching or apparent low-frequency noise in
transport.  In the well studied case of Random Telegraph Signals
(RTS) in Si Metal-Oxide-Semiconductor Field Effect Transistors
(MOSFET), electrostatic measurements reveal that the switching is
most likely caused by the fluctuating charge of interface
defects\cite{ralls}.  The charge fluctuations on the defects lead
to fluctuations in the scattering potential experienced by the
carriers. In small devices this leads to bi- or multi-valued
fluctuating conductance\cite{ralls}, while in devices with a large
number of defects it results in $1/f$-like frequency dependent
noise\cite{inv_f}.

At low temperatures, the charge fluctuation of the defects are
mediated by quantum mechanical electron tunneling between the
localized defect states and the itinerant states.  The signature
of the quantum-mechanical tunneling is the temperature-independent
switching rate at low temperatures\cite{r1}.  In this problem,
we must also consider the Coulomb interaction, which in fact
makes RTS observable.   Tunneling in
the presence of strong electron correlations has been of active
interest as it exhibits a number of peculiar features. One of
these features is the Fermi-edge singularity in the resonant
impurity tunneling. The singular dependence of the tunnel rate on
the energy of the impurity electron relative to the Fermi energy
of conduction electrons was first suggested by Matveev and Larkin
~\cite{r40}, whose theory was based on the X-ray
emission/adsorbtion problem. The latter was extensively studied in
numerous works, see Ref.~\cite{r5} and references therein, which
have shown that the transition rate is proportional to
$E^\alpha$, where $E$ is excess energy relative to the
absorption/emission threshold.  Nozieres and De Dominicis
~\cite{r6} have demonstrated that the power $\alpha$ is related to
the scattering phase shift of the conduction electrons at the core
hole potential and it has two essential contributions: a negative
one due to exciton-like physics\cite{r5}, i.e. an attraction between
the core hole and excited electron, and a positive one due to the
orthogonality catastrophe\cite{pwa}, i.e. an adjustment of the
Fermi sea to the appearance of the core hole potential.  For a
single channel model, i.e., in case of spinless Fermi sea
electrons interacting with point-like core hole potential,
Nozieres and De Dominicis obtained that
\begin{eqnarray}
\alpha = - (2/\pi){\rm tan}^{-1}(\nu V) + [{\rm tan}^{-1}(\nu V)/\pi]^2\, ,\label{a0}
\end{eqnarray}
where ${\rm tan}^{-1}(\nu V)$ is the scattering phase shift, $\nu$
is the density of states of the conduction electrons and $V$ is
the Coulomb interaction strength. In the case of a local defect
tunnel-coupled to a conduction channel, the same model applies after
making an association between the X-ray electrical field coupling
strength ${\cal E} d$ and the tunneling matrix element $\Delta$,
and defining the energy $E$ as the difference between the localized
level position and the Fermi level.  Then, one finds
that the tunnel rate between the impurity and the Fermi sea
is~\cite{r40}:
\begin{eqnarray}
\gamma_{\rm in/out} = \gamma_0 \Theta(\pm E_0)|E_0 \tau_e|^\alpha\, ,\label{a02}
\end{eqnarray}
\noindent where $\gamma_0 = 2\pi \nu\Delta^2$, $\Theta$ is a step
function, $\tau_e$ is the cutoff of the order  of the conduction
electron Fermi energy and $\alpha$ is given by Eq.~(\ref{a0}).

The singular behavior predicted by Eq.~(\ref{a02}) has been
observed experimentally in the resonant impurity tunneling in Si
MOSFETs~\cite{r1}, with a negative exponent $\alpha$.  There is,
however, a serious discrepancy between the simple theory
prediction, Eq.~(\ref{a02}), and experiments~\cite{r1,r02}  that
has to do with the prefactor $\gamma_0$. In particular, in Si
MOSFETs the tunneling is found to occur on a rather slow
timescale, typically from milliseconds to seconds.
\begin{figure}[htbp]
\begin{center}
\includegraphics[width = 3.1 in]{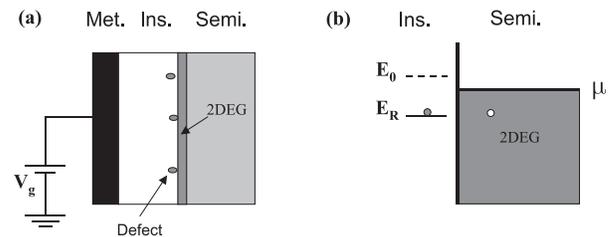}
\vspace{0.5cm} \caption{(a) Schematics of MOSFETs; (b) Energy diagram for the impurity level and
the conducting electrons. Coupling to optical phonons shifts the bare position of the level.}
\label{fig:setup}
\end{center}
\end{figure}
\vspace{-0.5cm} Given such tunneling timescales one might have guessed that the slow tunneling rate
is due to a large distance, on the order of 15-20~\AA, between the impurities and the
insulator-semiconductor interface. However, the distance between the resonant impurity and the
2-Dimensional Electron Gas (2DEG) in the MOSFET conducting channel can be determined via a quite
simple experimental procedure based on the monitoring of the defect average occupancy as a function
of the gate voltage, which reveals that the separations between the impurities and the 2DEG is
typically of the order of 1-2~\AA~\cite{r7}. A naive estimate suggests that for such width of the
tunnel barrier (of height of several eV) the impurity electron dwell time must be in the range of
pico -to nanoseconds, which is by many orders of magnitude smaller than the observed dwell time. In
the present work we argue that this apparent contradiction can be resolved if one takes into
account strong electron-optical phonon coupling in the Si MOSFET insulating layer, which
drastically slows down the defect charge dynamics. To this end, we solve the problem that
simultaneously includes the effects of electron-electron and electron-phonon interactions. We find
that at energies $E$ small compared to the optical phonon frequency one recovers the singular
behavior of Ref.~\cite{r40} but with a renormalized ``bare'' tunneling rate $\tilde\gamma_0 $,
while at sufficiently large energies (for $E > E_p$, where $E_p$ is the polaronic shift defined
below) the renormalization vanishes.  We make numerical estimates which indicate that the
renormalization effect can be very strong in a Si MOSFET.

A schematic diagram of MOSFET is presented in Fig. 1(a). The impurity is located near the interface
in the insulating layer.  The Hamiltonian of the system includes the electronic part $H_0$ and the
electron-phonon interaction $H_{\rm ph}$.  The electronic part is
\begin{eqnarray} H_0 = \sum_k E_k c^{\dag}_k c_k + E_0 d^{\dag} d + V d^{\dag} d\sum_{k,k^{\prime}}
c^{\dag}_k c_{k^\prime}&& \nonumber\\
+\Delta\sum_{k}(c^{\dag}_k d + d^{\dag} c_k)\, .~~~~~~~~~~~~~~\label{a01}
\end{eqnarray}
\noindent Here $c^\dag_k(c_k)$ and $d^\dag(d)$ are creation(annihilation) operators of spinless
electrons in the Fermi sea and at the impurity respectively (we will discuss the role of spin
degeneracy later on), the third term describes Coulomb coupling of the impurity electron and the
Fermi sea electrons, and the last term represents tunneling between the impurity and conduction
electrons. We assume that the tunneling matrix element $\Delta$ is independent of the single
particle states in the Fermi sea and that the impurity potential is point-like. We also assume that
the density of states ($\nu$) of Fermi sea electrons is constant and set the chemical potential
($\mu$) in the Fermi sea to zero in the following calculations.

In general, electron-phonon interaction in bulk semiconductors is
insignificant.  However, in the insulating oxide layers of
MOSFETs, the polar structure of the material favors strong
electron-phonon coupling.  We model this local electron-phonon
coupling by a Holstein-type Hamiltonian
\begin{eqnarray}
H_{\rm ph} =g\, d^{\dag} d\, (a^\dag + a) + \omega_0 a^{\dag} a\, ,\label{a03}
\end{eqnarray}
where $a^{\dag}$($a$) are creation(annihilation)  operators for a local optical phonon of frequency
$\omega_0$ coupled to the occupation number at the impurity. The coupling constant $g$ will be
estimated later~\cite{r6}.

We now evaluate the defect electon tunnelnig rate at $T = 0$ by
calculating the matrix element $\langle 0 |e^{-itH} |0 \rangle$,
which is the probability amplitude for the impurity state to
remain empty. Here $H = H_0 + H_{\rm ph}$ and the ``ground'' state
$|0\rangle$ corresponds to the trap being empty, and the phonon
and the conduction electrons in the ground state, i.e., $|0
\rangle = |0\rangle_{\rm d} \otimes |0\rangle_{\rm ph}
\otimes|0\rangle_{\rm el}$. We will carry out the calculation in
Euclidean time by setting $it = \beta$. In what follows $\hbar$
and electron charge $e$ are set to unity unless stated otherwise.

We expand ${\cal Z} = \langle 0 |e^{-\beta H} |0 \rangle$ in terms of the tunneling term in the
Hamiltonian $H$ as
\begin{eqnarray}
{\cal Z}=\sum_N\int d^{2N}[\beta]\,\langle 0|{\cal T}e^{-\beta H^\prime}
H_T(\beta_{2N})\,.\,.\,.\,H_T(\beta_{1}) |0 \rangle ,\label{a5}
\end{eqnarray}
where we have introduced a shorthand notation $\int {\cal T}
d^{2N}[\beta] = \int_0^\beta
d\beta_{2N}\int_0^{\beta_{2N}}d\beta_{2N-1}...\int_0^{\beta_2}
d\beta_{1}$, ${\cal T}$ is the time ordering operator, and
$N=0,1,...\ $. In Eq.~(\ref{a5}) $H_T(\beta) = \exp{(\beta
H^\prime)} H_T \exp{(-\beta H^\prime)}$, where $H_T
=\Delta\sum_{k}(c^{\dag}_k d + d^{\dag} c_k)$ and $H^\prime$
contains all the remaining terms of the Hamiltonian $H$. Note that
$H^\prime$ is diagonal in $d^{\dag} d$ and can be decomposed into
two commuting parts, $H_{\rm el}$ and $H_{\rm ph}$, that involve
conduction electrons and phonons respectively.

The conduction electrons and optical phonons can be integrated out by realizing that a term with given $N$
in the sum in Eq.~(\ref{a5}) corresponds to trap site being occupied for
periods of time $\beta_{2i-1}< \beta < \beta_{2i}$, $i\leq N$, and being empty otherwise, see Fig.
2(a). One can formally write that term in the sum in Eq.~(\ref{a5}) as
\begin{eqnarray}
\Delta^{2N} \langle 0|{\cal T} \prod_{j=1}^{N}\ \sum_{k_{2j},k_{2j-1}}c^{\dag}_{k_{2j}}(\beta_{2j})
c_{k_{2j-1}}(\beta_{2j-1})\times&& \nonumber\\
{\rm e}^{-\int_0^\beta H_{\rm el}[n_d(\beta^\prime)]d\beta^\prime}|0\rangle_{\rm el} \ \langle
0|{\cal T}{\rm e}^{-\int_0^\beta H_{\rm ph}[n_d(\beta^\prime)]d\beta^\prime}|0\rangle_{\rm ph} \,
.\label{a06}
\end{eqnarray}
\noindent Here the explicitly time dependent Hamiltonians $H_{\rm el}[n_d(\beta)]$ and $H_{\rm
ph}[n_d(\beta)]$ correspond to the first three terms in Eq.~(\ref{a01}) and to Eq.~(\ref{a03})
respectively, with the replacement $d^\dag d \rightarrow n_d(\beta)$, where $n_d(\beta)$ is the
occupation number of the trap as a function of time,
\begin{eqnarray}
n_d(\beta) = \sum_{j=1}^N \left[\Theta(\beta-\beta_{2j-1})- \Theta(\beta-\beta_{2j})\right]\, .
\label{a54}
\end{eqnarray}
The phonon-dependent matrix in Eq.~(\ref{a06}) can be evaluated by a standard method of either
cummulant expansion or by path integral techniques as a ground -to ground state amplitude of a
harmonic oscillator subjected to an external force $n_d(\beta)$. By using Eq.~(\ref{a54}) this
matrix element is
\begin{eqnarray}
{\rm e}^{-GN}{\rm e}^{E_p\sum_j (-1)^j\beta_j}\exp{-G\sum_{j>k} (-1)^{j-k}{\rm
e}^{-\omega_0(\beta_j - \beta_k)}}\, , \label{a55}
\end{eqnarray}
\noindent where we have defined the polaronic shift $E_p = g^2/\omega_0$ and $G=E_p/\omega_0$. The
first exponent in Eq.~(\ref{a55}) is due to the diagonal terms ($j=k$) in the sum over $j,\,k$.

The electronic matrix element in Eq.~(\ref{a06}), which is essentially the $N$-particle propagator
for the conduction electrons, can be evaluated in the large $\beta$ limit by the technique
developed by Nozieres and De Dominics \cite{r6}, and by Anderson and Yuval~\cite{r8}. The technique
relies on the flat density of states assumption for the Fermi sea electrons (which is the case for
the 2DEG in the conduction channel) and is based on the theory of singular integral equations. It
allows one to obtain an asymptotically exact expression for the $N$-particle Green's function in
Eq.~(\ref{a06}) in the limit $|\beta_{2j}-\beta_{2j-1}| \gg \tau_e$. By adjusting the result of
Refs.~\cite{r6,r8} for the electronic Green's function in  Eq.~(\ref{a06}), and substituting Eq.
~(\ref{a55}) in Eqs.~(\ref{a5},\ref{a06}) we obtain
\begin{eqnarray}
{\cal Z} = \sum_N ({\gamma_0{\rm e}^{-G}\over 2\pi\tau_e})^N\int d^{2N}[\beta]{\rm e}^{-E_R
\sum_j(-1)^j\beta_j}\times
&& \nonumber\\
\exp{\sum_{j>k}(-1)^{j-k}[K\ln{({\beta_j-\beta_k \over\tau_e})} - G{\rm e}^{-\omega_0(\beta_j -
\beta_k)}]}\, . \label{a11}
\end{eqnarray}
\noindent In Eq.~(\ref{a11}) we have defined $E_R = E_0 - E_p$,
$\gamma_0 = 2\pi\nu\Delta^2$, and $K = 1+\alpha$. Thus, due to the
coupling to the phonon mode the resonant level shifts downwards as
shown in Fig. 1(b).
\begin{figure}[htbp]
\begin{center}
\includegraphics[width = 3.1 in]{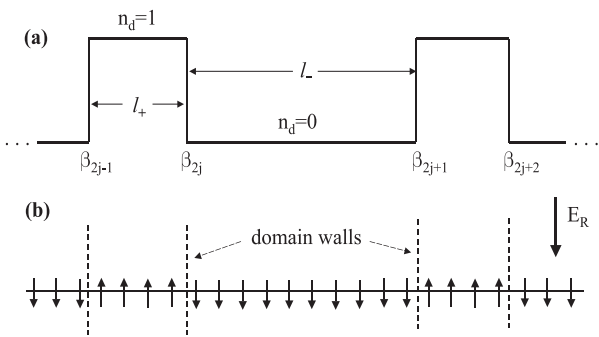}
\vspace{0.5cm} \caption{(a) Typical ``trajectory'' for the trap
occupation number as a function of time; (b) Spin domains in the
equivalent Ising model}
\end{center}
\end{figure}

To proceed, we write Eq.~(\ref{a11}) as
the $T = 1$ partition function of a one-dimensional ferromagnetic
Ising model, where a spin up or down corresponds to an empty or
occupied impurity respectively, as shown in Fig.~2(b).
The spacing in imaginary time
between successive spins equals $\tau_e$.
This Ising model has a nearest neighbor coupling term
$-(1/2)[\ln{(2\pi/\gamma_0\tau_e)} + G] s_i s_{i+1}$ ($s_i =\pm
1$), and a long range interaction $-(1/2)[K\,(i-j)^{-2} +
(G\omega_0^2\tau_e^2)\exp(-\omega_0\tau_e |i-j|)] s_i s_j$.
The renormalized position of the resonant level $E_R$
plays the role of an external magnetic field. In the weak long-range
interaction regime relevant here, $K < 1$, the Ising system is in
a paramagnetic state composed of ferromagnetic domains whose size
is controlled by the short range interaction.  A simple estimate
for the mean separation between the domain walls at zero external
field ($E_R = 0$) yields $l\approx\exp{(G)}/\gamma_0\gg\tau_e$.
The external field ($E_R\neq 0$) forces spins to align along its
direction, thus shrinking the domains opposing the field relative
to the ones aligned with the field. The ratio of the corresponding
domain sizes can be estimated by evaluating the average
magnetization $\langle s\rangle$ per spin.  For
$E_R\gg\gamma_0\exp{(-G/2)}$ one gets $1-\langle s\rangle \sim
\gamma_0\exp{(-G)}/(\tau_e E_R^2)$ and therefore the average size
of the domains with spins along the field exceeds that of the
domains of spins opposing the field at least by a factor of
$\eta=\tau_e E_R^2/[\gamma_0\exp{(-G)}]\gg 1$. (The long range
terms will favor even more polarized state as they provide
additional ferromagnetic coupling.)  As a result the interaction
between domain walls from different domains is negligible provided
$\omega_0, E_R \gg \gamma_0\exp{[-E_p/\omega_0]}$.

The above arguments allow for a significant simplification of the
transition matrix element $\cal Z$ in Eq.~(\ref{a11}) in the limit
of small $\gamma_0\tau_e\exp{(-G)}$. Retaining only those terms in
the last exponent of Eq.~(\ref{a11}) that couple $\beta_{2j}$ and
$\beta_{2j-1}$ we obtain that ${\cal Z} =$
\begin{eqnarray}
\sum_N \left({\gamma_0\over 2\pi\tau_e}\right)^N{\rm e}^{-GN}\int d^{2N}[\beta]\,
\prod_{j=1}^N\phi\left(\beta_{2j}-\beta_{2j-1}\right) \, , \label{a12}
\end{eqnarray}
\noindent where $\phi(\beta) = \exp{[-E_R\beta
-K\ln{(\beta/\tau_e)}+ G{\rm e}^{-\omega_0\beta}]}$. Our
approximation is similar to the non-interacting ``blip''
approximation used in the solution of the spin-boson problem
\cite{r12} and therefore the same method of Laplace transform can
be used to evaluate the sum in Eq.~(\ref{a12}). Upon Laplace
transform $\tilde{\cal Z} = \int_0^\infty d\beta
e^{-\beta\lambda}{\cal Z}(\beta)$, which converts the $N$'s term
in the sum into $(\Delta {\rm e}^{-G})^{2N}
(1/\lambda)^{N+1}[\tilde\phi(\lambda)]^N$, where
$\tilde\phi(\lambda)=\int_0^\infty d\beta
e^{-\beta\lambda}\phi(\beta)$. Re-summation in Eq.~(\ref{a12})
then yields $\tilde{\cal Z} = [\lambda - (\gamma_0/2\pi\tau_e){\rm
e}^{-G}\tilde\phi(\lambda)]^{-1}$. The tunnel rate $\gamma$ can be
associated with twice the imaginary part of the pole of
$\tilde{\cal Z}$. In the limit $\gamma_0\tau_e{\rm e}^{-G}\ll
E_R$, the rate $\gamma$ of tunneling into the impurity is given by
\begin{eqnarray}
{\gamma_0{\rm e}^{-G}\over 2\pi\tau_e}{\rm Im}\int\exp{\left[-E_R\beta -
K\ln{({\beta\over\tau_e})}+ G{\rm e}^{-\omega_0\beta}\right]}d\beta \, , \label{a130}
\end{eqnarray}
\noindent  In order to evaluate the imaginary part of the above integral an analytical continuation must be used. We follow prescription of Ref.~\cite{r20}. The procedure is
straightforward: at the saddle point $\beta_0$ (given by equation $-E_R-(K/\beta_0)= G\exp{(-\omega_0\beta_0)}$) we deform the contour of integration into the complex plane in the
direction of steepest descent (along the $+i$ direction in our case). As a result the integral picks up a convergent imaginary part. The numerical result is presented in Fig. 3. In the
limiting cases of small and large $E_R$ simple asymptotic expressions for the tunnel rate into the impurity are possible:
\begin{mathletters}
\label{a13}
\begin{eqnarray}
&&\gamma_{\rm in} = {\gamma_0{\rm
e}^{-E_p\over\omega_0}\Theta(-E_R)\over\tilde{\Gamma}(1+\alpha)}|E_R\tau_e|^\alpha\, ,\ |E_R|\ll
\omega_0\,;
\label{a13a}\\
&&\gamma_{\rm in} = {\gamma_0\Theta(-E_R)\over
\tilde{\Gamma}(1+\alpha)}[(E_R-E_p)\tau_e]^\alpha\, ,\ |E_R|>
E_p\,. \label{a13b}
\end{eqnarray}
\end{mathletters}
In Eqs.~(\ref{a13}) $\tilde{\Gamma}(z) = (2\pi/z)^{1/2} z^z {\rm
e}^{-z}$. For $E_R\ll E_p$ the rate is suppressed by
$\exp{(-E_p/\omega_0)}$.  This suppression occurs in case of {\em
elastic} tunneling, when no real phonon modes are excited, and
essentially represents the overlap of the phonon ground state wave
function for occupied and empty defect states.~\cite{r12}.  The
exponent $\alpha$ is due the  X-ray singularity as in
Eq.~(\ref{a02}). For $E_R\sim E_p$ the exponential renormalization
disappears due to opening of additional {\em inelastic} tunneling
channels with excited phonons in the the final state.

So far we have not included the effects on spin degeneracy.
However, based on the close analogy between our results
Eqs.~(\ref{a130},\ref{a13}) and the X-ray absorption ~\cite{r6}
problem, we argue that the main effect of the spin degeneracy is
to change the second term in the expression for the exponent
$\alpha$ in Eq.~(\ref{a0}) by a factor of 2.

\begin{figure}[htbp]
\begin{center}
\includegraphics[width = 3.1 in]{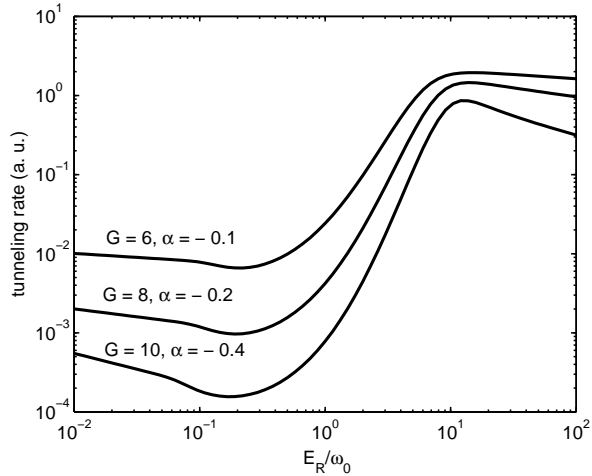}
\vspace{0.5cm} \caption{Dependence of the tunnel rate on the bias}
\end{center}
\end{figure}
For the interface impurity states in a Si MOSFET the coupling
constant turns out to be sufficiently strong. In order to estimate
the order of magnitude of the polaronic shift that enters
Eqs.~(\ref{a13}) ($E_p = g^2/\omega_0$), we take Fr\"{o}hlich
coupling constants of an electron to an optical phonon mode with a
wavevector $k$~\cite{r9}, $g_k = \alpha\langle\Psi_d|
e^{ikr}|\Psi_d\rangle/(V^{1/2}|k|)$. Here $V$ is crystal volume
and $\alpha^2 = 2\pi e^2 \omega_0(\epsilon_\infty^{-1} -
\epsilon_0^{-1})$, where $\epsilon_\infty$ and $\epsilon_0$ are
high frequency and static dielectric constants, $e$ is electron
charge and $\Psi_d$ is the wavefunction of the impurity electron.
Assuming that the latter is a hydrogen-like ground state with
effective Bohr radius $a_d$, one finds that $\langle\Psi_d|
e^{ikr}|\Psi_d\rangle = [1+(a_d k)^2/4]^{-2}$. Using the
equivalence between electron-phonon coupling in our single-mode
model and Fr\"{o}hlich's multi-mode Hamiltonian ~\cite{r60}, that
is $g^2=\sum_k g^2_k$, we obtain that $E_p = 5
e^2/(16a_d)(\epsilon_\infty^{-1} - \epsilon_0^{-1})$. For a deep
impurity level in ${\rm SiO_2}$, $a_d\sim 1 \AA$,
$\epsilon_\infty\simeq 2$, $\epsilon_0\simeq 4$, we get $E_p
\simeq 1\,{\rm eV}$. The frequency of the optical phonons near the
${\rm SiO_2-Si}$ interface is of the order $60\,{\rm meV}$, which
yields the renormalization of the tunnel rate in Eq.~(\ref{a13a})
by a factor of $\exp{(-16)} \simeq 10^{-7}$. While the detailed
quantitative theory warrants further study, we believe that our
model gives a qualitative explanation for the tunneling slowdown
at ${\rm Si-SiO_2}$ interfaces.

The authors are grateful to H.W. Jiang, M. Weissman, J. Zaanen, S. Lyon, S. Trugman, S.
Charkavarty, and Y. Imry for useful discussions. This work was supported by the US DOE. D.M. was
supported, in part, by the US NSF, grant DMR-0121146.


\begin{references}
\vspace{-15mm}
\bibitem{ralls} K.S. Ralls {\em et al.}, Phys. Rev. Lett. {\bf
52}, 228 (1984).
\bibitem{inv_f} M. J. Uren, D. J. Day, and M. J. Kirton, Appl.
Phys. Lett. {\bf 47}, 1195 (1985).
\bibitem{r1} D. H. Cobden and B. A. Muzykantskii, Phys. Rev. Lett. {\bf 75}, 4274 (1995); D. H.
Cobden, M. J. Uren, and M. Pepper, {\it ibid.}, {\bf 71}, 4230
(1993).
\bibitem{r40} K. A. Matveev and A. I. Larkin, Phys. Rev. B {\bf 46}, 15337 (1992).
\bibitem{r5} G. D. Mahan, {\it Many-Particle Physics} (Plenum Press, New York, 1981).
\bibitem{r6} P. Nozieres and C. T. De Dominics, Phys. Rev. {\bf 178}, 1097 (1969).
\bibitem{pwa} P.W. Anderson, Phys. Rev. Lett. {\bf 18}, 249
(1967).
\bibitem{r02} K. S. Ralls and R. A. Buhrman, Phys. Rev. Lett. {\bf 60}, 2434
(1988); M.-H. Tsai, H. Muto, and T.P. Ma, Appl. Phys. Lett. {\bf 61}, 1691 (1992).
\bibitem{r3} M. Xiao, I. Martin, and H. W. Jiang, Phys. Rev. Lett. {\bf 91}, 078301 (2003).
\bibitem{r60} The phonon Hamiltonian in Eq.~(\ref{a03}) should, in general, include miltiple phonon modes.
However, as long as the optical phonons have roughly the same frequencies, the Hamiltonain with a
single mode is equivalent to the multiple mode Hamiltonain with the appropriate choice of the
coupling constant $g$.
\bibitem{r7} G. Ghibaudo and T. Boutchacha, Microelectronics
Reliablity {\bf 42}, 573 (2002).
\bibitem{r8} P. W. Anderson and G. Yuval, Phys. Rev. Lett. {\bf 23}, 89 (1969).
\bibitem{r9} H. Fr\"{o}hlich, Adv. Phys. {\bf 3}, 325 (1954).
\bibitem{r12} A. J. Leggett {\it et al}., Rev. Mod. Phys. {\bf 59}, 1 (1987).
\bibitem{r20} S. Coleman, {\it Principles of Symmetry}, (Cambridge University Press, 1985).
\end{references}
\end{document}